\documentclass[
prl,
preprint,
showpacs,
showkeys,
aps,
]{revtex4}
\usepackage{amsmath,amssymb,amsfonts}
\usepackage{graphicx}

\begin{document}
\title{Nonresonant contributions to energy transfer through micron-size gaps between neighboring nanostructures}

\author{Luciano C. Lapas$^1${\email[E-mail:~]{luciano.lapas@unila.edu.br}}, Agust\'{i}n P\'{e}rez-Madrid$^2${\email[E-mail:~]{agustiperezmadrid@ub.edu}}, and J.~Miguel Rub\'{\i}$^2${\email[E-mail:~]{mrubi@ub.edu}}}

\affiliation{
$^1$Universidade Federal da Integra\c{c}\~{a}o Latino-Americana, Caixa Postal 2067, 85867-970 Foz do Igua\c{c}u, Brazil\\
$^2$Departament de F\'{\i}sica Fonamental, Facultat de F\'{\i}sica, Universitat de Barcelona, Av. Diagonal 647, 08028 Barcelona, Spain
}


\begin{abstract}
Current theoretical approaches to the analysis of radiative heat exchange at the nanoscale are based on Rytov's stochastic electrodynamics. However, this approach falls short in the description of microscale energy transfer since it overlooks non-resonant contributions arising from the coupling between different modes of relaxation in the material. We show that the phonon density of states given through a log-normal distribution accounts for such mode-coupling and leads to a general expression for the heat transfer coefficient which includes non-resonant contributions. This expression fits the existing experimental results with remarkable accuracy.  Thus, our theory goes beyond stochastic electrodynamics and offers an overall explanation of energy transfer through micrometric gaps regardless of geometrical configurations.
\end{abstract}

\maketitle

\emph{1. Introduction.}--- Heat transfer between two objects at nanometric scales maintained at different temperatures was first studied following a stochastic or fluctuational electrodynamics formalism established by Rytov {\em et al.}~\cite{Rytov59} and Polder and van Hove~\cite{Polder71}. More recent studies have taken into account Casimir and van der Waals forces~\cite{Lifshitz55}, evanescent waves~\cite{Pendry99,Rousseau09}, surface phonon polaritons~\cite{Shen09} or have used Landauer-like reformulation~\cite{Biehs08}, to cite some examples. All these approaches have something in common: electrostatic interactions, linear response regime and, consequently, the fluctuation-dissipation theorem (FDT). Thus, an approach based on the contributions of fluctuating dipole effects seems to be the heart of a considerably simplified treatment of energy transfer at the nanoscale~\cite{Shen09,Rousseau09,Domingues05}. These approaches include a wide range of phenomena in which the energy transfer between molecules is dominated by dipole-dipole interactions, also known as F\"{o}rster energy transfer~\cite{Forster48}. Nonetheless, as two nanostructures, thermalized at different temperatures, come closer to each other, the distribution of charges and currents becomes asymmetric and therefore, defies description in terms of dipolar interactions. Hence, it is clear that one must bear in mind higher orders effects beyond the dipole-dipole~\cite{Madrid08}.  In this sense, stochastic electrodynamics falls short in the whole description of microscale energy transfer since it overlooks non-resonant contributions to the system response which violates the FDT mentioned above ~\cite{Madrid09}. These non-resonant contributions come from the phonon-photon interaction  and arise in the coupling between the different modes of relaxation in the material. In addition, some of the previous studies have apparently called into question the validity of the Derjaguin's approximation~\cite{Shen08,Shen09,Rousseau09} since an imbalance of action and reaction can occur in out-of-equilibrium situations~\cite{Buenzli09}.

In this letter we present a mesoscopic treatment of this process in order to shed light on previous controversies. The assumption of a log-normal as the phonon density of states, which takes into account such mode-coupling,  allows us to obtain a general expression for the heat-transfer coefficient including resonant and non-resonant contributions. Our theory goes beyond stochastic electrodynamics and offers an overall explanation of energy transfer through micrometric gaps regardless of geometrical configurations, fitting the existing experimental results with a high degree of accuracy.

\emph{2. Mesoscopic non-equilibrium thermodynamics.}--- A thermodynamic description entails the formulation of the second law. This can be carried out by means of the Gibbs entropy postulate~\cite{deGroot84}, 
\begin{equation}
S(t)=-\frac{k_B}{V}\int n (p,t)\ln \frac{n (p,t)}{n _{eq}(p)}d\mathbf{p} +S_{eq}. \label{entropy1}
\end{equation}
Here $n (p,t)$ is the momentum distribution of photons, $d\mathbf{p}$ is the volume element in momentum space and $V$ the volume. This gives us the non-equilibrium entropy of the photon gas plus the bath, where $S_{eq}$ is the equilibrium entropy and $n _{eq}$ the equilibrium distribution. Entropy is produced due to irreversible processes and its production is always positive, in accordance with the second law. Irreversible processes in non-equilibrium systems are described by means of currents, thermodynamic forces and the entropy production rate.
The presence of a momentum distribution gradient $\partial n (p,t)/\partial p$ yields a current $J\left( p,t\right)$ which satisfies the relation $J\left( p,t\right) =-D \partial n (p,t)/\partial p$, derived from the entropy production~\cite{deGroot84}. This relation accounts for the diffusion of probability between different regions of the phase space. Here, $D(\sim p^2$/time) is a kinetic coefficient, the diffusion coefficient in momentum space. 

\begin{figure}[!h]
\includegraphics[scale=2]{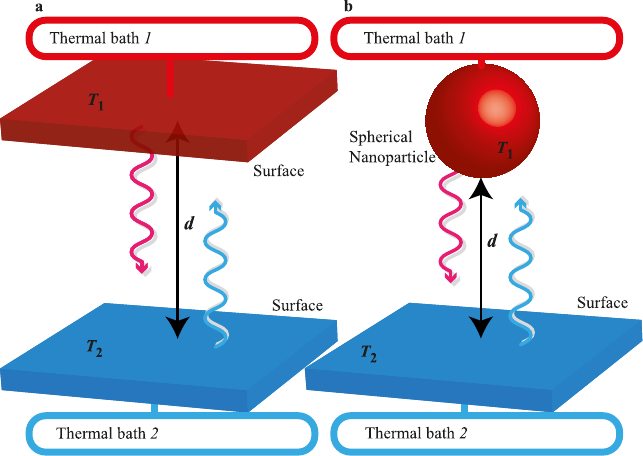}
\caption{(color online). Schematic model. Schematic diagram of the radiation exchanged between two materials maintained at different temperatures, $T_1$ and $T_2$, separated by a distance $d$. \textbf{a}, Two parallel surfaces. 
\textbf{b}, Sphere-surface system.}
\label{fig:scheme}
\end{figure}

This scheme can be applied to the description of the radiative heat transfer between two surfaces at temperatures $T_{1}$ (hot) and $T_{2}$ (cold), respectively, separated by a distance $d$, Fig. \ref{fig:scheme}a. Since in the gap between both surfaces there are only hot and cold photons interacting, it is plausible to assume that the average current in this gap results from the superposition of hot and cold currents, $J_{H}(t)$ and $J_{C}(t)$, respectively,
\begin{equation}
J(p,t)=J_{C}(t)\delta \left( p-p_{C}\right)+J_{H}(t)\delta \left( p-p_{H}\right). \label{current}
\end{equation}
These currents are related to the net rate of emission of photons on both surfaces and thus must be proportional to the density of photons.
When the characteristic length scales are comparable to the wavelength of thermal radiation, $\lambda _{T}= ch /k_{B}T$, the diffusion coefficient might depend on frequency. Here, $h$ is the Planck's constant. For length scales $d\gg \lambda _{T}$ ({\it i.e.} low frequencies) the diffusion coefficient must be constant (the wave character disappears). On the other hand, when $d\lesssim \lambda _{T}$ ({\em i.e.} high frequencies) system-size effects become important and the diffusion coefficient should depend on the ratio $\lambda _{T}/d$ or equivalently on the frequency~\cite{Madrid09}, $ D(\omega )\equiv\left(h^2c/V\right)\hat{D}(\omega )$. If the temperatures are kept constant, hot and cold photons will reach equilibrium with their respective baths. Hence, it follows from the integration of the previous general expression of the diffusion current with the help of Eq. (\ref{current}) that a stationary value $J_{st}(\omega)=D(\omega )\times \lbrack n(\omega ,T_{1})-n(\omega ,T_{2})]$, is established. Here, $n(\omega ,T)/V=2N(\omega ,T)/h^{3}$, with $N(\omega ,T)$ being the averaged number of quasi-particles in a elementary cell of volume $h^{3}$ of the phase-space given by the Planck's distribution~\cite{Planck91}, $N(\omega ,T)=1/\left[ \exp \left( \hslash \omega /kT\right) -1\right] $. 

Since each photon carries an energy $\hbar \omega $, the heat flow $Q$ follows from the sum of all the contributions as $Q=\left( 1/h^2\right)\int \hslash \omega J_{st}(\omega )d\mathbf{p}$, where $\mathbf{p}=\left( \hslash \omega/c\right) \boldsymbol\Omega_{p}$. For the proposed case we assume that the density of vibrational states is achieved through the use of a log-normal distribution, which corresponds to 
\begin{equation}
\hat{D}(\omega )=\frac{c^2k^2_c}{ \sigma \omega _{0}\sqrt{2\pi }}\exp \left\{ -\left[ \frac{ \ln (\omega /\omega _{0})}{\sigma \sqrt{2} } \right]^{2}\right\} \delta(\omega-\omega _{R}) \text{,} \label{eq:D} 
\end{equation}
with $k_{c}=1/\lambda _{T}$, a law which has been successful in a somewhat similar problem analyzed previously~\cite{Denisov90}. It may be considered that the log-normal results from the existence of a hierarchy of relaxation mechanisms in the material~\cite{Madrid09}. This law differs from the Debye approximation for the density of states $\omega ^{2}/\pi ^{2}c^{3}$ related to purely vibrational modes and is a characteristic of disordered systems to which dynamics is mainly due to slow relaxing modes, as for example glasses~\cite{Debenedetti98}.

Therefore, this contribution comes clearly from the excited phonons in the material due to the photon-phonon interaction. Here, $\omega _{0}$ (characteristic frequency), and $\sigma $ (standard deviation) are two fitting parameters, and $\omega_{R}\propto 1/d$ is a resonance frequency. In the blackbody radiation limit (far-field), $\hat{D}(\omega )=1/4$ for which we obtain $Q=\sigma _{B}\left( T_{1}^{4}-T_{2}^{4}\right) $ which constitutes the Stefan-Boltzmann non-equilibrium law, \ with $\sigma_{B}=\pi ^{2}k_{B}^{4}/60\hbar ^{3}c^{2}$ being the Stefan constant. Nevertheless, there must be a multi-scale transition regime between the low-frequency and the high-frequency regimes for which $Q$ contains contributions $\sum_n c_n/d^n$, where $c_n$ corresponds to different length scales taking into account the material emissivity~\cite{Rousseau09}, evanescent wave contributions~\cite{Pendry99,Shen09}, Casimir and electrostatic forces~\cite{Shen09,Lifshitz55,Rousseau09}, and so on. This multi-scale regime is accounted for by assuming that $\hat{D}(\omega )=\sum_n \lambda_n \omega^n \delta(\omega - \omega_R)$, where $\lambda_n\propto c_n$. As we have shown in Ref.~\cite{Madrid08}, the modification of the distribution of charges or currents due to proximity between objects induces multipolar interactions of an order higher than the dipolar interaction. This hierarchy of multipoles introduces a concomitant hierarchy of length scales. Accordingly, the multiscale regime mentioned above which includes the different contributions and effects referred to in the literature, clearly corresponds to a multipolar field expansion.

In experiments, one usually measures the heat transfer coefficient $H\left( \omega _{R},T_{0}\right) \equiv Q/\left( T_{1}-T_{2}\right) $, with $T_{0}=\left( T_{1}+T_{2}\right) /2$, for which we obtain 
\begin{gather}
H\left( \omega _{R},T_{0}\right) =4\sigma _{B}T_{0}^{3}\left(\lambda_0+\lambda_1\omega _{R}+\lambda_2\omega _{R}^{2}\right) + \notag \\
\kappa \omega _{R}^{2}\exp \left\{ -\left[ \frac{\ln (\omega _{R}/\omega
_{0})}{\sigma \sqrt{2}}\right] ^{2}\right\} \left( \frac{\hslash \omega
_{R}/k_{B}T_{0}}{\sinh \left( \hslash \omega _{R}/2k_{B}T_{0}\right) }%
\right) ^{2}. \label{eq:transfer}
\end{gather}
where $\kappa \equiv k^2_{c}k_{B}/[(2\pi )^{5/2}\sigma \omega _{0}]$. Here, we have taken into account terms up to the second order, involving far-field, electrostatic and out-of-equilibrium contributions.

\begin{figure}[!h]
\includegraphics[scale=1]{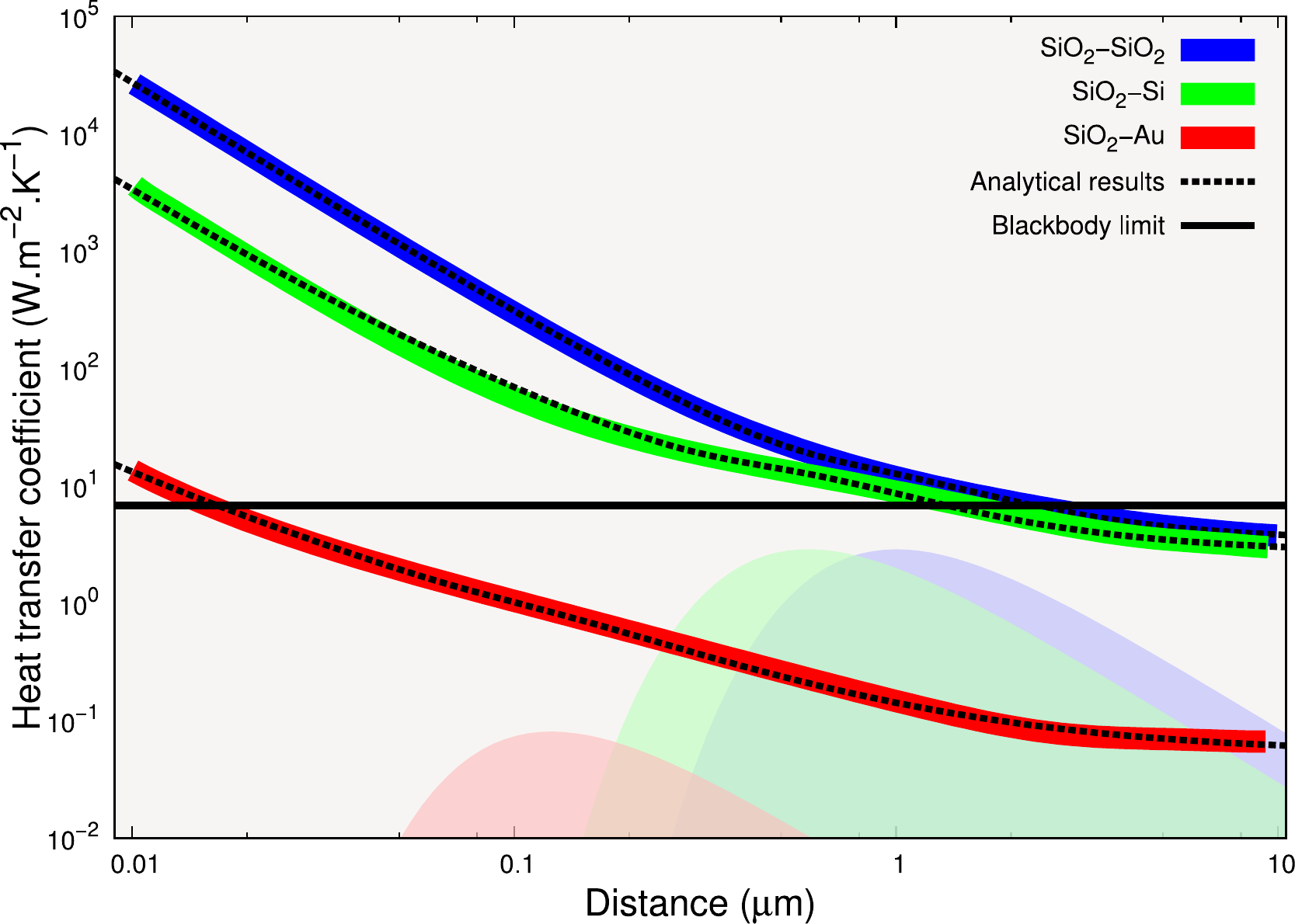}
\caption{(color online). Radiative heat transfer coefficients between two parallel plates as a function of the gap obtained by Shen \textit{et al.}~\cite{Shen09}. The black solid line is the limit of thermal radiation predicted by the usual blackbody radiation law, where the heat flux is calculated from Stefan-Boltzmann law at an average temperature $T=300$ K. The black dashed lines show the analytical result obtained from Eq. (\ref{eq:transfer}) by adjusting all parameters to the experimental data. The filled curves correspond to the non-Debye behavior for each of the material configurations, glass-glass, glass-silicon and glass-gold.}
\label{fig:plate-plate}
\end{figure}

\emph{3. Results and discussion.}--- In order to test Eq. (\ref{eq:transfer}), we compare it with the near-field thermal radiation measurements between two parallel plates made of three different polar dielectric material combinations ($SiO_{2}-SiO_{2}$, $SiO_{2}-Si$, and $SiO_{2}-Au$)~\cite{Hu08,Shen09}. Figure \ref{fig:plate-plate} shows a strong enhancement of the heat conductance for short distances, exceeding to the blackbody limit. Additionally, the contribution containing the log-normal in Eq. (\ref{eq:transfer}) has been plotted in all the cases as shaded areas into the Fig. \ref{fig:plate-plate}. 

Let us analyze the thermal conductance between a microsphere and a plate. In a previous discussion~\cite{Shen09}, the authors pointed out that Derjaguin's approximation~\cite{Derjaguin56} does not hold for near-field radiative heat transfer. In contrast, other authors~\cite{Rousseau09} demonstrated that this approach is in agreement with the data in the range 2.5 $\mu$m to 30 nm. Equation (\ref{eq:transfer}) which avoids the use of Derjaguin's approximation, provides a general way which sheds light on the theory of radiative heat transfer at the nanoscale. Assuming the scheme in Fig. \ref{fig:scheme}b, the distance $d$ between both surfaces depends on the curvature measured through the local radius $r$, as described in Ref.~\cite{Rousseau09}, $\omega_R=\omega _{R}(d(r))$. Here, we use $\tilde{d}=d+b+R-\sqrt{R^2-r^2}$, where $b$ represents a roughness parameter~\cite{Rousseau09}. Hence, a more accurate measurement of the thermal conductance is given by the surface integral of Eq. (\ref{eq:transfer}), $\bar{H}= \int_0^R H\left(\omega_{R},T_{0}\right) 2\pi r dr$. 

\begin{figure}[!h]
\includegraphics[scale=1]{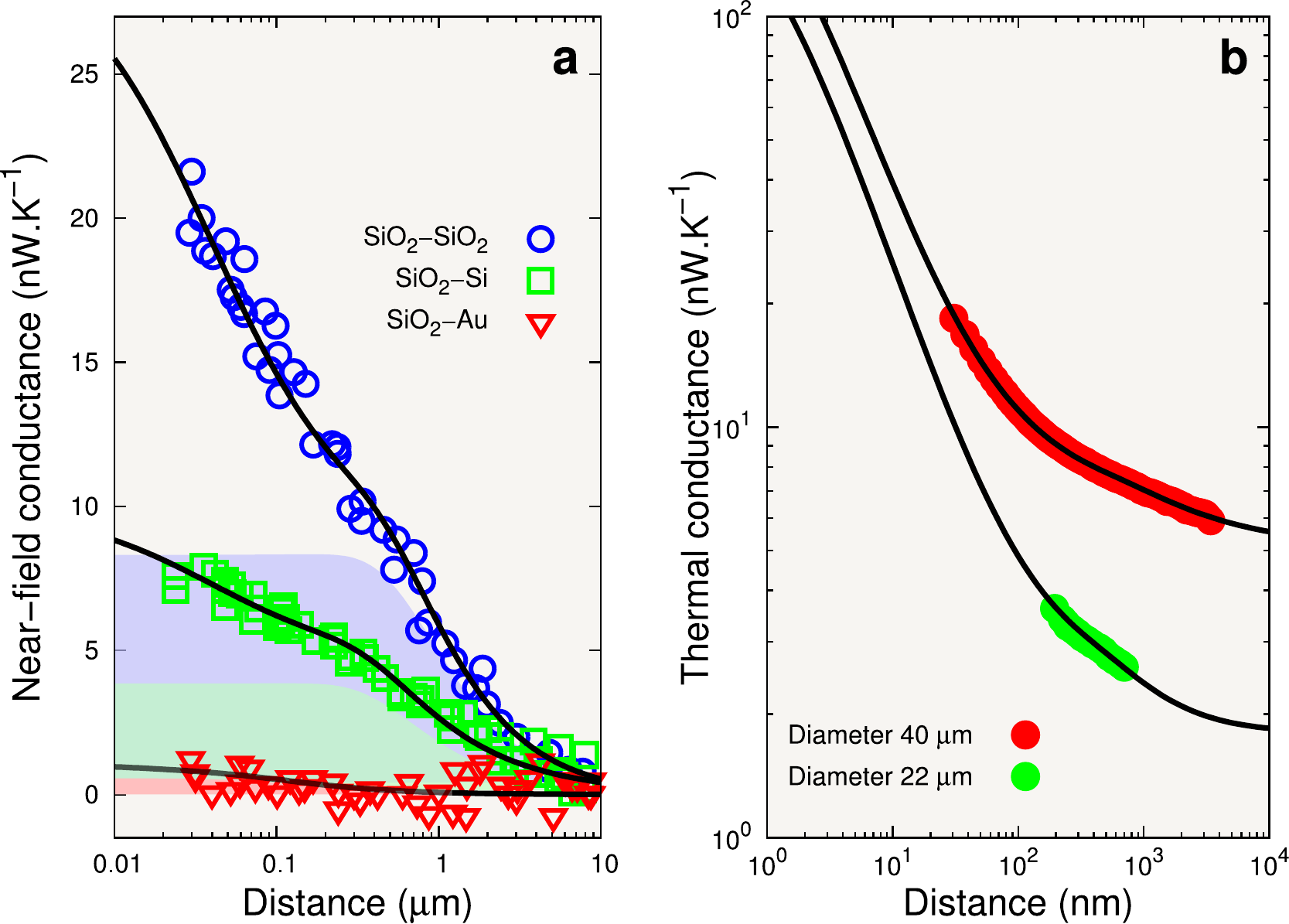}
\caption{(color online). Thermal conductance between sphere and surface as function of the gap. \textbf{a}, Experimental data from the near-field radiation measurements between a silicon dioxide (glass) sphere and different substrate materials (glass, silicon, and gold) obtained by Shen \textit{et al.} \protect~\cite{Shen09}. The shaded areas correspond to non-Debye regime represented by the log-normal contribution in Eq. (\ref{eq:transfer}): blue for glass-glass, green for glass-silicon, and red for glass-gold material combinations. \textbf{b}, The red and green dots represent experimental data obtained by Rousseau \textit{et al.} \protect~\cite{Rousseau09} with a sphere of different radius.}
\label{fig:sphere-plate}
\end{figure}

From this, we calculate numerically the surface average by adjusting the parameters mentioned above to the experimental data obtained in the Refs.~\cite{Rousseau09,Shen09}. In Fig. \ref{fig:sphere-plate}a we plotted the near-field conductance fitting Eq. (\ref{eq:transfer}) to the data~\cite{Shen09}. Shaded areas reveal the influence of glassiness on the thermal conductance, which is quite similar to the contribution of non-resonant evanescent waves to near-field radiation~\cite{Shen09,Pendry99}. As a further check of the theory, we compare measurements with a microsphere of different radius~\cite{Rousseau09}, Fig. \ref{fig:sphere-plate}b. In the range 30 nm to 2.5 $\mu$m, the roughness parameter provides a slight deviation from $1/d$ asymptotic behavior, in such a way that the transition between the far and near-field occurs because of the non-Debye contribution, in contrast with previous conclusions~\cite{Rousseau09}. Therefore, glassiness is a crucial ingredient in the near-field radiative heat exchange.

Additional information about the value of the parameters for fitting Eq. (\ref{eq:transfer}) to the experimental data can be found in the supplementary material.

\emph{4. Conclusion.}--- In summary, we have evaluated thermal conductance in a wide range of length scales, from the far-field to the near-field, giving a thermokinetic description of several experiments involving heat radiation through a very narrow gap. Although near-field radiative transfer is a highly complex phenomenon, we have been able to provide a unified and highly accurate explanation of heat exchange processes at the nanoscale as well as a description of the transition between the far-field and the near-field regimes. Since the experiments examined may involve a great variety of nanostructures, our theory possesses a wide scope of applications. The general methodology presented here may also be used in the study of other heat exchange processes such as those occurring in phonon systems and in the analysis of thermal contributions to Casimir forces.

\subsection*{Acknowledgments}
This work was supported by the Brazilian National Council for Scientific and Technological Development (CNPq) under Grant No. 309094/2010-0 and MICINN of the Spanish Government under Grant No. FIS2008-04386.

\end{document}